\documentclass[aps,prev,twocolumn,preprintnumbers,floatfix,nofootinbib]{revtex4-1}
\pdfoutput=1

\newcommand{\be}{\begin{equation}}
\newcommand{\ee}{\end{equation}}
\newcommand{\bea}{\begin{eqnarray}}
\newcommand{\eea}{\end{eqnarray}}

\usepackage{mathrsfs}
\usepackage{amsmath, amsthm, amssymb}
\usepackage{color}
\usepackage{epstopdf}
\usepackage[pdftex]{graphicx}
\usepackage{amssymb}
\usepackage{verbatim}
\usepackage{hyperref}
\usepackage{enumerate}
\usepackage{float}
\usepackage[caption = false]{subfig}
\usepackage[normalem]{ulem}

\newcommand{\Tscans}{6 months}

\newcommand{\VTI}{$1$ cm$^3$}

\newcommand{\mrange}{0.7 to 3.5 meV}
\newcommand{\BiFeSe}{(Bi$_{1-x}$Fe$_x$)$_2$Se$_3$}

\begin{document}

\title{A Proposal to Detect Dark Matter Using Axionic Topological Antiferromagnets}

\author{David J. E. Marsh$^{a}$}
\email{david.marsh@uni-goettingen.de}
\author{Kin Chung Fong$^{b}$}
\author{Erik W. Lentz$^{a}$}
\author{Libor \v{S}mejkal$^{c,d,e}$}
\author{Mazhar N. Ali$^{f}$}

\vspace{1cm}
\affiliation{${}^a$ Institut f\"{u}r Astrophysik, Georg-August Universit\"{a}t, Friedrich-Hund-Platz 1, D-37077 G\"{o}ttingen, Germany}
\affiliation{${}^b$ Raytheon BBN Technologies, Quantum Engineering and Computing, Cambridge, Massachusetts 02138, USA}
\affiliation{${}^c$Institut f\"ur Physik, Johannes Gutenberg Universit\"at Mainz, D-55099 Mainz, Germany}
\affiliation{${}^d$Institute of Physics, Academy of Sciences of the Czech Republic, Cukrovarnick\'{a} 10, 162 53 Praha 6 Czech Republic}
\affiliation{${}^e$Faculty of Mathematics and Physics, Charles University in Prague, Ke Karlovu 3, 121 16 Prague 2, Czech Republic}
\affiliation{${}^f$ Max Planck Institute of Microstructure Physics, Weinberg 2, 06120 Halle (Saale), Germany}

\begin{abstract}

Antiferromagnetically doped topological insulators (A-TI) are among the candidates to host dynamical axion fields and axion-polaritons; weakly interacting quasiparticles that are analogous to the dark axion, a long sought after candidate dark matter particle. Here we demonstrate that using the axion quasiparticle antiferromagnetic resonance in A-TI's in conjunction with low-noise methods of detecting THz photons presents a viable route to detect axion dark matter with mass \mrange, a range currently inaccessible to other dark matter detection experiments and proposals. The benefits of this method at high frequency are the tunability of the resonance with applied magnetic field, and the use of A-TI samples with volumes much larger than 1 mm$^3$.

\end{abstract}

\maketitle
Astrophysical and cosmological observations provide strong evidence for the existence of non-baryonic dark matter (DM)~\cite{2005PhR...405..279B,planck_2015_params,2015NatPh..11..245I,pdg}. Among possible candidates are dark axions (DA)~\cite{1983PhLB..120..133A,1983PhLB..120..137D,1983PhLB..120..127P,2010RvMP...82..557K,2012PDU.....1..116R,2015ARNPS..65..485G,Marsh:2015xka}, hypothetical particles ~\cite{pecceiquinn1977,weinberg1978,wilczek1978} suggested to solve the charge-parity ($\mathcal{CP}$) problem in quantum chromodynamics (QCD)~\cite{Afach:2015sja}. Searching for the DA is challenging due to its weak coupling to ordinary matter (e.g. photons). For DA masses $m_a\lesssim 0.2\text{ eV}$ the local DA field, $\theta_D$, can be described as a classical coherent state. The local DM density is then $\rho_{DM} = |\theta_D(t)|^2 m_a^2 f_a^2/2$, where $m_a$ and $f_a$ are the unknown axion mass and ``decay constant'', and the measured value is $\rho_{\rm DM}\approx 0.4 \text{ GeV cm}^{-3}$~\cite{2015NatPh..11..245I}. The DA field oscillates in time, with a frequency dominated by the rest energy, $m_a c^2$, and an intrinsic width set by the galactic velocity dispersion, $\sigma_v\approx 230 {\rm km\, s}^{-1}\Rightarrow \Delta\omega_a/\omega_a=\sigma_v^2/c^2\approx 10^{-6}$. The QCD axion mass can be computed in chiral perturbation theory or on the lattice, and is given by $m_a = 0.6 \text{ meV}(10^{10}\text{ GeV}/f_a)$~\cite{weinberg1978,wilczek1978,Borsanyi:2016ksw} (we use units $\hbar=c=1$ if not stated otherwise). The central frequency is $\nu = 0.25 (m_a/\text{ meV})\text{ THz}$. 

Only one DM search, the Axion Dark Matter eXperiment (ADMX)~\cite{2010PhRvL.104d1301A,Du:2018uak}, has made a significant constraint on the QCD axion parameter space predicted by the Kim-Shifman-Vainshtein-Zhakarov (KSVZ)~\cite{1979PhRvL..43..103K,1980NuPhB.166..493S} and Dine-Fischler-Srednicki-Zhitnitsky (DFSZ)~\cite{Zhitnitsky:1980tq,1981PhLB..104..199D} models. The QCD axion mass can span $10^{-12}\lesssim m_a\lesssim 10^{-2}\text{ eV}$, satisfying astrophysical constraints on the couplings~\cite{2008LNP...741...51R,2014PhRvL.113s1302A} with $f_a$ less than the Planck scale. There are hints, however, pointing to the meV range, particularly for DFSZ-type models~\cite{Kawasaki:2014sqa,2012MNRAS.424.2792C}. Furthermore, constraints from the CERN Axion Solar Telescope (CAST)~\cite{2014PhRvL.112i1302A} and the Any Light Particle Search~\cite{2010PhLB..689..149E}, combined with the prediction of KSVZ and DFSZ models provide a target range for the axion-photon coupling $10^{-13}\text{ GeV}^{-1}< g_\gamma< 10^{-10}\text{ GeV}^{-1}$ for $m_a= 1\text{ meV}$. In the meV range, axion DM searches also overlap with searches for spin-dependent forces~\cite{2014PhRvL.113p1801A,Kim:2017yen}.

\begin{figure}
\includegraphics[width=1\columnwidth]{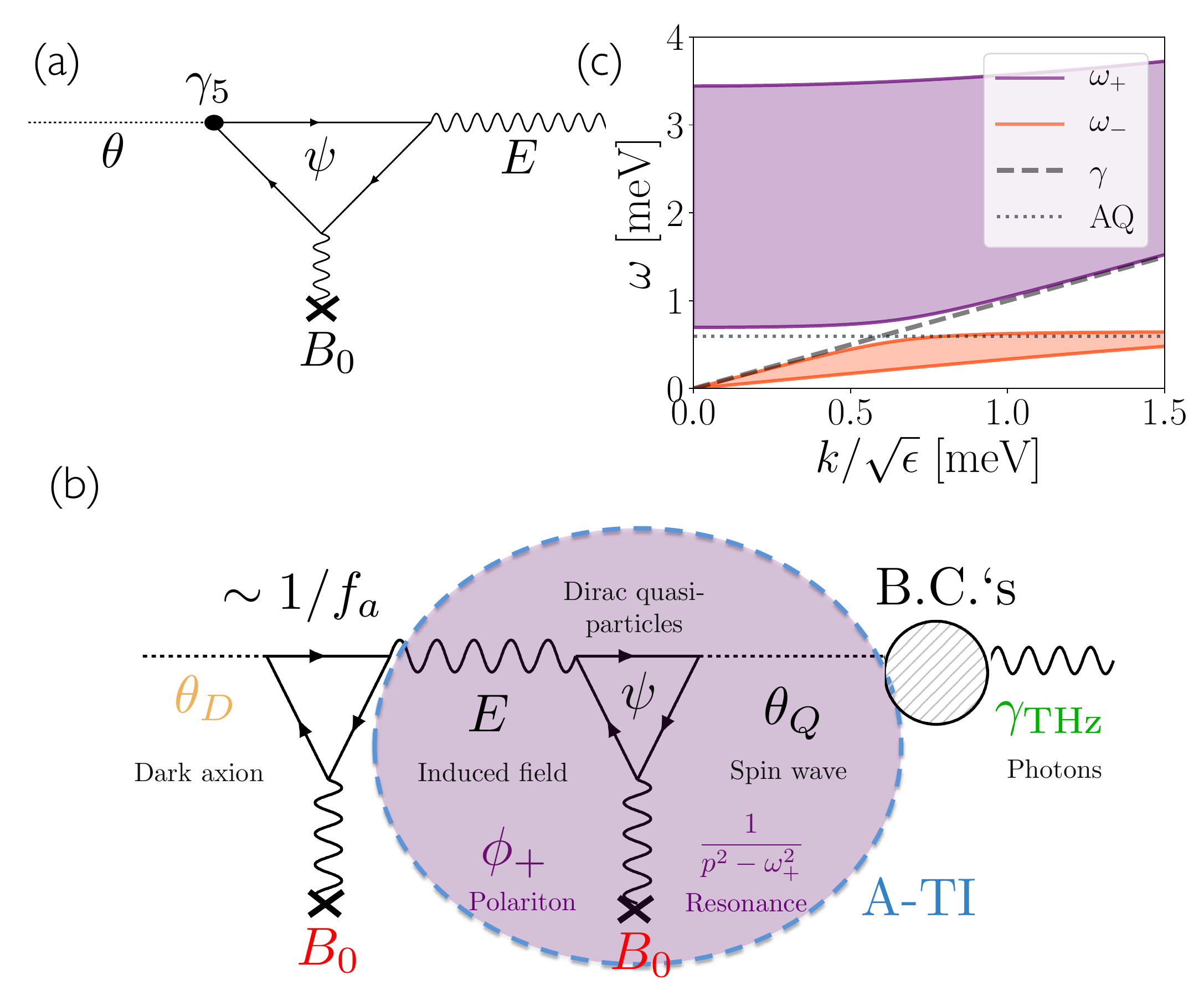}
\caption{(a) The chiral anomaly~\cite{Adler:1969gk,Bell:1969ts}. $\theta$ is a pseudoscalar chirally coupled to charged Dirac fermions, $\psi$. With applied $B_0$, $\theta$ mixes with $E$ leading to the existence of axion-polaritons, $\phi_\pm$, in the case of $\theta_Q$ and the production of photons in the case of $\theta_D$. (b) Resonant enhancement of DA-photon conversion. Coloured text refers to Fig. 3. Inside the A-TI the DA couples to the mixed states $\phi_\pm$ shown in the shaded circle. Conversion is resonantly enhanced when $p^2=\omega_a^2=\omega_+(k,B_0)^2$, represented by the polariton propagator. At the A-TI dielectric boundary, polaritons convert to propagating photons, due to boundary conditions (B.C.'s)~\cite{Millar:2016cjp} represented here by the vertex. (c) The axion-polariton dispersion relation for $\omega_\pm(k,B_0)$~\cite{2010NatPh...6..284L}. Scanning the applied $B_0$ field tunes $\omega_+(k=0)$ in the range \mrange\, and scans the resonance.}
 
\label{fig:feynmans}
\end{figure}

The power output from the axion-induced electric field (Fig.~\ref{fig:feynmans}a) is: 
\be
P_0 = \frac{1}{2} E_0^2 V_{\rm eff} \omega_a =  g_\gamma^2 B_0^2 \frac{\rho_{\rm DM}}{m_a^2} V_{\rm eff} \omega_a \, .
\label{eqn:p0}
\ee
Taking the effective volume $V_{\rm eff}\approx (2\pi/m_a)^3$ from the vacuum dispersion relation, $m_a=1\text{ meV}$, $g_\gamma=10^{-10}\text{ GeV}^{-1}$, and $B_0=1\text{ T}$, gives $P_0=10^{-27}\text{ W}$. Detecting the axion requires amplifying this power. Methods to amplify the signal include resonance in a microwave cavity~\cite{2010PhRvL.104d1301A,Du:2018uak,Zhong:2018rsr,McAllister:2017lkb}, ferromagnetic resonance~\cite{Crescini:2018qrz}, coherent enhancement~\cite{TheMADMAXWorkingGroup:2016hpc}, and many others~\cite{2014PhRvL.112m1301S,2016PhRvL.117n1801K,2017PDU....18...67M,2017PDU....15..135B,2014PhRvX...4b1030B,Crescini:2018qrz,Abel:2017rtm,Stadnik:2013raa}. In particular, a resonant cavity haloscope method at mm wavelengths enhances the signal by the quality factor, $Q$, but suffers from small effective volume, since the resonance requires $V\sim (2\pi/\omega_a)^3$. The highest frequency operating cavity haloscope is ORGAN, at 0.1 meV~\cite{McAllister:2017lkb}, while the MADMAX dielectric haloscope projects maximum frequencies of 0.5 meV~\cite{TheMADMAXWorkingGroup:2016hpc}. In this letter we propose an alternative method that combines THz resonant enhancement and volume increase facilitated by axion-photon conversion inside a topological axion insulator antiferromagnet. To estimate the signal strength we use antiferromagnetically Fe-doped Bi$_2$Se$_3$ as a realistic possibility, demonstrating that it fits our general requirements through symmetry analogy to the Fu-Kane-Mele-Hubbard model of antiferromagnetic diamond.

Axionic degrees of freedom are predicted to materialise as quasiparticles, $\theta_Q$, in magnetically doped topological insulators (TIs)~\cite{2010NatPh...6..284L}, Cr$_2$O$_3$ ($\theta_{Q}=\pi/36$)~\cite{Essin2009,Malashevich2012}, $\alpha$-Fe$_2$O$_3$ \cite{2011PhRvL.106l6403W} with a corundum structure, spinels~\cite{Wan2012} and magnetic TI heterostructures~\cite{Wang2016DAQ}. The signatures of the topological magnetoelectric effect, a.k.a. \textit{static} axion electrodynamics, were recently reported as quantized magneto-optical effects in TIs~\cite{Wu2016,dziom2017,Okada2016}, and quantized magneto and electrical resistance changes in artificial antiferromagnetic heterostructures of magnetically doped TIs~\cite{Mogi2017,Xiao2018,Grauer2017}.  Finally, dynamical axion quasiparticles (AQ) in the form of magnetic fluctuations were predicted in magnetically doped TIs (MTI)~\cite{2010NatPh...6..284L}, spin-orbit coupled Mott insulators~\cite{Sekine2016}, and in MTI superlattices~\cite{Wang2016DAQ}. 

We propose to use AQs in antiferromagnetically doped TIs (A-TI) to detect DAs. The conversion process of DAs to visible photons is shown in Fig.~\ref{fig:feynmans}(b). Antiferromagnets provide the correct THz frequency range owing to the resonance frequency exchange enhancement $\omega\sim\sqrt{(2H_{E}+H_{A})H_{A}}$ ($H_{E},H_{A}$ are exchange and anisotropy fields respectively). Inside the A-TI, AQs mix with the electric field $E$ and generate (quasi-particle) axion-polaritons (AP, see Fig.~\ref{fig:feynmans}a,b), $\phi_\pm$ \cite{2010NatPh...6..284L} (see also Ref.~\cite{Tercas:2018gxv}). When $\omega_\pm(k,B_0)=\omega_a$, the conversion process is resonantly enhanced by $Q=\omega/\Gamma$, where $\Gamma$ is the polariton damping (width). If the allowed values of $k$ are restricted by the geometry, then the lowest value of $k\sim1/L$ can facilitate resonant conversion of DAs to APs in volumes much larger than $(c/\text{THz})^3$. The combination of $Q$ and $V$ allows the signal power to be greatly enhanced compared to $P_0$. The APs convert into propagating photons due to the boundary conditions (B.C.'s)~\cite{Millar:2016cjp}, and can be detected. As we will now show, this detection strategy gives access to a unique part of DA parameter space.

We begin by defining an axionic field $\theta_i$ by the coupling to the electromagnetic Chern-Simons (CS) term generated by the loop Fig.~\ref{fig:feynmans}(a):
\begin{equation}
S_{\rm CS} =\sum_{i={\rm D,Q}} \frac{\alpha}{\pi}C_i \int {\rm d}^4 x \theta_i \mathbf{E}\cdot \mathbf{B} \, ,
\label{action}
\end{equation}
where $\mathbf{E}, \mathbf{B}$ are electric and magnetic fields. $\theta_D$, is a pseudoscalar pseudo-Goldstone boson with a non-vanishing electromagnetic chiral anomaly~\cite{pecceiquinn1977,weinberg1978,wilczek1978,Adler:1969gk,Bell:1969ts}. The coupling $C_i$ is dimensionless: the dimensionful axion-photon coupling is defined by $g_\gamma = C_i\alpha/2\pi f_i$. $C_D$ is a model-dependent constant taking the values $C_{\rm KSVZ}=-1.92$ and $C_{\rm DFSZ}=0.75$, and $f_i=f_a$. For the AQ, we define $C_{Q}=1$.  

Other DA couplings to ordinary matter~\cite{Srednicki:1985xd,2013PhRvD..88c5023G,2015ARNPS..65..485G} could also affect the A-TI. Nuclear spin couplings lead to resonance at the Larmor frequency, which with $B\lesssim 20$ T gives $\nu\lesssim 100\text{ MHz}$~\cite{2013PhRvD..88c5023G,2014PhRvX...4b1030B}, far below the DA frequency at 1 meV. The axion-electron coupling induces DA absorption in Dirac semi-metals~\cite{2018PhRvD..97a5004H}. The parameter space with significant absorption, however, is excluded by astrophysical constraints. Thus we neglect the direct nuclear and electron DA couplings.
 
The criteria for generating AQs in condensed matter as suggested by Wilczek are~\cite{1987PhRvL..58.1799W}: (i) effective action in the form of Eq.~\eqref{action} (ii) realization of the Dirac equation for electrons and (iii) tuneable Dirac masses. 

Criterion (i) can be met in general in magnetoelectric materials with nonzero diagonal components of the magnetoelectric polarisability tensor $\alpha_{ij}=\left( \frac{\partial M_{j}}{\partial E_{i}}\right)_{\textbf{B}=0}=\left( \frac{\partial P_{i}}{\partial B_{j}}\right)_{\textbf{E}=0}$, where $M,P$ are magnetization and electric polarisation.  
Since $\theta_Q$ is odd under spatial inversion $\cal{P}$ and time reversal $\cal{T}$, and the physical observables $\sim e^{iS/\hbar}$ (where $S$ is the action) are defined modulo $2\pi$, the CS term can be nonzero in (a) magnetoelectric matetials with a magnetic point groups with broken $\cal{P}$, and broken $\cal{T}$ where $\theta$ is nonquantized, (b) $\theta=\pi$ can be taken as a defining property of $\mathcal{T}$-invariant TIs~\cite{Essin2009,Wu2016}.

Criterion (ii) can be realised in Dirac quasiparticle materials such as TIs where the simultaneous presence of $\cal{P}$, and $\cal{T}$ symmetries protects the Kramers double degeneracy of the bulk Dirac bands, while at the surfaces realise $\cal{T}$ protected 2D Dirac quasiparticle helical states~\cite{Hasan2010}. To satisfy (iii) and generate dynamical axion fields, gradients of $\theta$ need also be generated dynamically, one possibility being magnetic fluctuations~\cite{2010NatPh...6..284L,Sekine2016,Wang2016DAQ}. In such a case, $\theta_Q$ is the pseudoscalar component of the spin wave.

\begin{figure}
\vspace{-0.2em}\includegraphics[width=1\columnwidth]{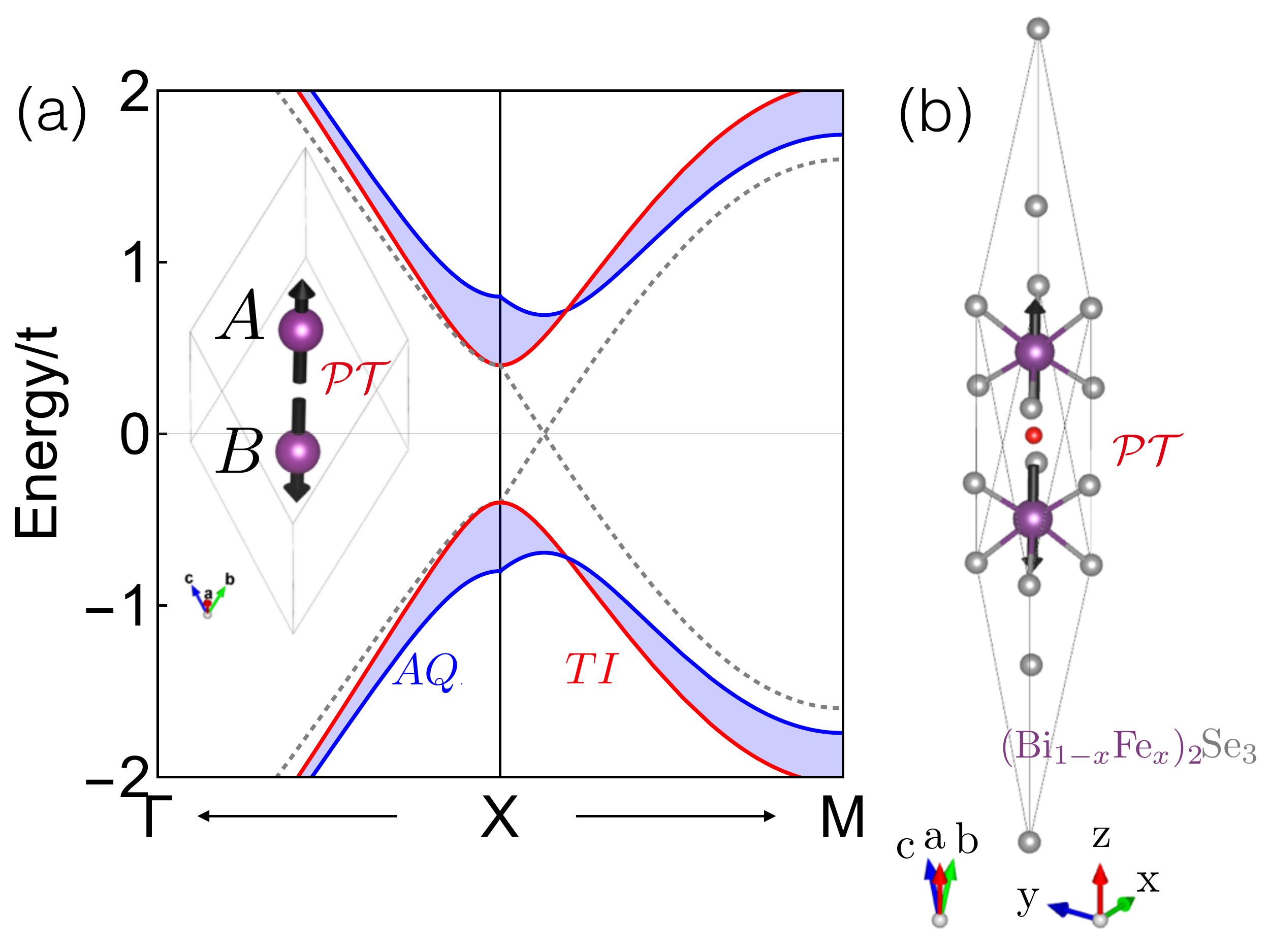}
\caption{Axion spin wave Dirac quasiparticle antiferromagnets. (a) Band structure of tuneable Dirac quasiparticles. TI: $\lambda=0.5, \delta t_{1}=0.4,Um=0$. AQ: $\lambda=0.5, \delta t_{1}=0.4,0<Um<0.25$. Inset:  Antiferromagnetic diamond lattice with marked $\mathcal{PT}$
symmetry. (b) Crystal of AF doped \BiFeSe\, exhibits the same magnetic point group symmetry as (a).}  
\label{fig:inverse_effect}
\end{figure}

To simultaneously satisfy all three criteria for AQs we identify Dirac quasiparticle antiferromagnets as suitable candidates~\cite{Smejkal2017,Smejkal2018,Sekine2014}. We consider a Dirac antiferromagnetic insulator with $\cal{P}$ and $\cal{T}$ symmetry broken and thus magnetoelectric point group, but importantly the combination $\mathcal{PT}$ preserved, with a generic electronic Dirac Hamiltonian $H(\textbf{k})=\sum_{i=1,..,5}A_{i}(\textbf{k})\gamma_{i}$, where $\gamma_{i}$ are Dirac matrices and $A_{i}(\textbf{k})$ parameterise the band structure. The antiferromagnetic coupling couples in proper basis choice to $\gamma_{5}$ in the Dirac Hamiltonian~\cite{2010NatPh...6..284L,Sekine2014}. 

As a particular realization we consider the antiferromagnetic Fu-Kane-Mele Hubbard~\cite{2008PhRvL.100i6407F} model on the bipartite (orbital degree of freedom $\tau$) diamond lattice with two spins per lattice site $\sigma$~\cite{Sekine2014,Rachel2018}. The antiferromagnetism breaks $\mathcal{T}$, and $\mathcal{P}$, but preserves $\mathcal{PT}$ as marked by the red ball in Fig.~\ref{fig:inverse_effect}(a) and thus preserves the form of the Dirac Hamiltonian. The Hamiltonian with a Hubbard term treated on a mean-field level reads:  
$
H=\lambda\left(\textbf{A}(\textbf{k})-\frac{U}{\lambda}\textbf{m}\right)\cdot \sigma \tau_{z}+t\text{Re}f(\textbf{k})\tau_{x}+t\text{Im}f(\textbf{k})\tau_{y},
$
where the nearest neighbour hopping on the diamond lattice (cf. Fig.~\ref{fig:inverse_effect},a) $f(\textbf{k})=\sum_{j=1,..,4}(t+\delta t_{j})e^{i\textbf{k}\cdot\textbf{d}_{j}}$ ($\textbf{d}_{j}$ being the four nearest neighbour vectors), $A_{x}(\textbf{k},Um_{x})=4 \sin \frac{k_{x}}{2}\left(\cos \frac{k_{y}}{2}-\cos \frac{k_{z}}{2}\right)$ plus cyclic permutations, $U$ is Hubbard correlation strength, $\lambda$ is strength of the spin-orbit coupling, and $\delta t_{j}$ represent the renormalization of the hopping due the deformation of the $AB$  bond. The AQ has a mean value given by  $\theta_Q=\frac{\pi}{2}[1+\text{sign}(\delta t_{1})]-\arctan (\frac{Um}{\delta t_{1}})$~\cite{Sekine2014}. It was shown that the fluctuations in the N\'{e}el order parameter $\textbf{L}\sim\textbf{m}_{A}-\textbf{m}_{B}$ (with axis along $z$) can be within approximation $\frac{U\vert\textbf{m}\vert}{\lambda}<< 1$ related to dynamical fluctuations of $\theta_{Q}$ ~\cite{2010NatPh...6..284L,Sekine2014}:
\begin{equation}
\delta \theta_{Q} \sim \frac{2}{3}\sum_{i=x,y,z}Um_{i} \, .
\label{dyna}
\end{equation}

The band structure of our model is shown in Fig.~\ref{fig:inverse_effect}(a) for a realistic range of effective exchange coupling $U\textbf{m}\sim 0 - 0.40$ and illustrates the tuning of the Dirac bands with a Dirac point shifted slightly off the $X$ ($\sim Um/2\lambda$) point due to the effect of antiferromagnetism. The AQ spin wave (SW)~\cite{1951PhRv...82..565K,RichardsAFR,1953AmJPh..21..250K} dispersion on the diamond lattice is 
\begin{equation}
\hbar\omega_{Q_{A}}\approx g\mu_{B} H_{0}\pm\sqrt{\left(8SJf(0)+g\mu_{B}H_{A}\right)^{2}-\left(8SJf(\textbf{q})\right)^{2}}, \label{SW}
\end{equation}
where $g\approx 1$ is the Land\'{e} factor, $\mu_B$ is the Bohr magneton, $H_{E}=8SJ$, and $\textbf{q}$ is the spin wave wave-vector. The AQ-SW tunes, in a first-order approximation, only the $z$-component of the $\textbf{L}\propto \mathbf{M}_A-\mathbf{M}_B$ order parameter~\cite{2010NatPh...6..284L,Sekine2014} (where $\mathbf{M}_{A,B}$ is the magentization on $A$, $B$ sublattices), which therefore tunes the Dirac mass as schematically illustrated by the shaded region in Fig.~\ref{fig:inverse_effect}(a).

No antiferromagnetic bulk dynamical axionic insulator has yet been identified in the lab. Remarkably, however, our model has exactly the same magnetic point group, $\overline{3}'1m'$, as the mean-field medium of Fe-doped Bi$_{\text{2}}$Se$_{\text{3}}$. This can be seen by deforming the face centred cubic primitive unit cell (Fig.~\ref{fig:inverse_effect},a) along the [111] direction to produce the rhombohedral unit cell of tetradymite Bi$_{\text{2}}$Se$_{\text{3}}$ (Fig.~\ref{fig:inverse_effect},b). It can be shown that the antiferromagnetism couples to the same $\gamma_{5}$ matrix as in our model~\cite{2010NatPh...6..284L}, and applying the Neumann principle gives axion-field favourable nonzero diagonal symmetric elements to $\alpha_{ij}$, and leads to the analogous expression for the AQ-SW field, Eq.~\eqref{dyna}.

The quadratic action for small fluctuations $|\theta_Q|<1$ is given by: 
\be
S_{\rm AQ} = \frac{f_Q^2}{2}\int {\rm d}^4x  \left[ \dot{\theta}_Q^2  - (v_{Q,i}\partial_i\theta_Q)^2 - m_Q^2\theta_Q^2\right] \, ,
\label{eqn:em_aa_system}
\ee
where $f_Q$ and $\mathbf{v}_Q$ are the SW stiffness and velocity. Scanning $\omega_\pm(B_0)$ (see Fig.~\ref{fig:feynmans},c) requires specifying $m_s(B_0)$ and  $f_Q(B_0)$. For \BiFeSe\, using Eq.~\ref{SW} with doping factor at 3.5\%~\cite{2013PhRvL.110m6601K}, exchange of 1 meV~\cite{2012PhRvL.109z6405Z} and anisotropy of 16 meV~\cite{2013PhRvB..88w5131Z}, the spin wave mass is $m_Q=[0.12 (B_0/2\text{ T})+0.6] \text{ meV}$.  From Ref.~\cite{2010NatPh...6..284L} we find $f_Q = 190\text{ eV}$ at $B_0=2\text{ T}$, and take $f_Q^2\propto 1/m_Q$ from the $\delta \mathbf{L}$ kinetic term~\cite{Wang2016DAQ}.

Including the usual Maxwell term, linearizing for small fluctuations in $E$ and $\theta_Q$ in the presence of an applied magnetic field, $B_0$, and external DA source, we find the system of equations derived from the action take the form (see also Refs.~\cite{1987PhRvL..58.1799W,Millar:2016cjp,Millar:2017eoc}): 
\begin{align}
\epsilon\ddot{\mathbf{E}}- \nabla^2\mathbf{E}+\frac{\alpha}{\pi}[\mathbf{B}_0\ddot{\theta}_{Q}-\nabla(\nabla\theta_Q\cdot \mathbf{B}_0)]& = \mathbf{A}\cos \omega_at \, , \nonumber\\
\ddot{\theta}_{Q}-v_Q^2\nabla^2\theta_{Q}+m_Q^2\theta_{Q}-\frac{\alpha }{4\pi^2 f_Q^2}\mathbf{B}_0 \cdot \mathbf{E} &=0 \, ,
\label{eqn:driven_system}
\end{align}
where $\epsilon=\epsilon_{r}\epsilon_{0}$ is the TI dielectric constant.

The driving term $\mathbf{A}=2 \mathbf{B}_0g_\gamma \sqrt{2 \rho_{\rm DM}}/m_a$ at leading order, and derives from Eq.~\eqref{action} taking the DA as an external source, with $\theta_D$ fixed by $\rho_{\rm DM}$. Neglecting the AQ dispersion compared to $\mathbf{E}$, we diagonalize Eq.~\eqref{eqn:driven_system}to find $\phi_\pm$ and $2\omega_\pm^2(k) = (k^2/\epsilon+m_Q^2+b^2)\pm\sqrt{(k^2/\epsilon+m_Q^2+b^2)^2-4k^2m_Q^2/\epsilon}$, where $b^2=\alpha^2B_0^2/4\pi^3\epsilon f_Q^2$~\cite{2010NatPh...6..284L} (see Fig~\ref{fig:feynmans}c), and $k$ is the Fourier conjugate of $x$. Dynamical AQs are required for the mixing: in the absence of derivatives, $\theta_Q$ and $E$ decouple in Eq.~\eqref{eqn:driven_system}. The presence of axion-polaritons can be verified using an inverse ``light shining through a wall''~\cite{1987PhRvL..59..759V} experiment (as described elsewhere~\cite{2010NatPh...6..284L}), which can measure $\omega_\pm(k,B_0)$ from the band gap. 

\begin{figure}
\vspace{-0.2em}
\includegraphics[width=1.\columnwidth]{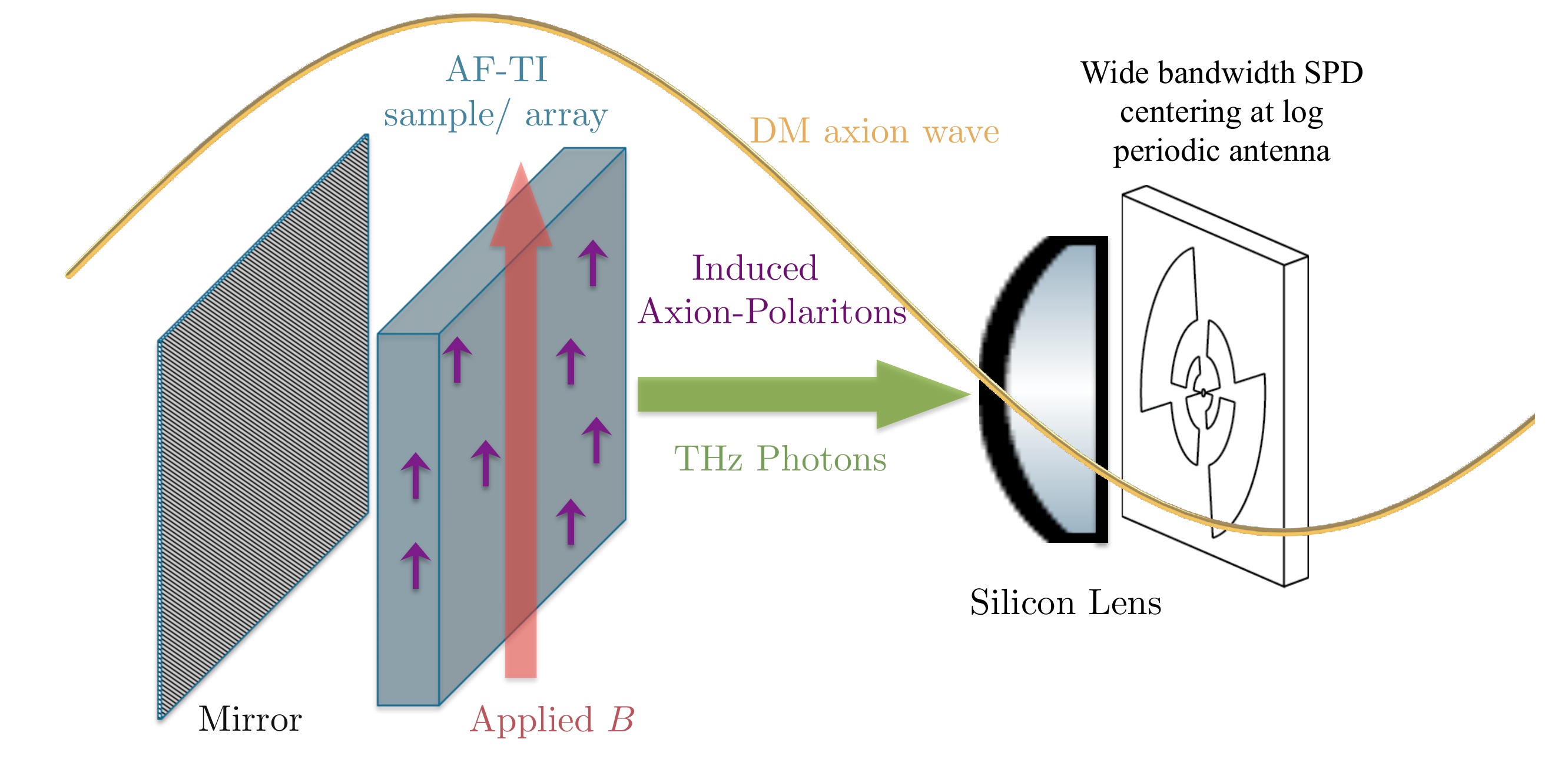}
\caption{Schematics of experimental concept. EM waves are emitted from all material surfaces perpendicular to $B_0$. A mirror and silicon lens focus THz photons from A-TI samples onto a single-photon detector located at the centre of a log periodic antenna.
}  
\label{fig:diagram}
\end{figure}

DA-driven polariton waves in the A-TI are a combination of $\mathbf{L}$, and $\mathbf{E}$. In the presence of $\mathcal{T}$ breaking, the A-TI surface states are gapped~\cite{2010RvMP...82.3045H,2010NatPh...6..284L}. The DA-induced surface polariton $E$-field thus leads to emission of photons from the surface of the A-TI, just like a dielectric haloscope, or dish antenna~\cite{2013JCAP...04..016H,TheMADMAXWorkingGroup:2016hpc}. If there is only one mode at a given $\omega$ then dielectric BC's are sufficient to compute the photon emission from polaritons at the boundary. We propose to detect the emitted photons by using a silicon lens to focus them onto a wide bandwidth single photon detector (SPD). A mirror placed behind the A-TI coherently enhances the forward emission~\cite{Millar:2016cjp}. The concept is illustrated in Fig.~\ref{fig:diagram}. 

The material boundary conditions and mirror restrict the allowed modes $k$ in the A-TI. The lowest lying mode, $k_0\sim 1/L$ (where $L$ is the system size), has the largest integrated volume, and we define $\omega_+(k_0,B_0):=\omega_+^0(B_0)$. The $E$-field power generated by resonant DA-photon conversion on resonance, $\omega^0_+(B_0)=\omega_a$, can be expressed as~\cite{1983PhRvL..51.1415S}, $P_{\rm signal} = (\omega/Q)\times (\text{energy stored})$: 
\be
P_{\rm signal} =\frac{1}{2}\kappa f_+ Q_{\rm sys} V_{\rm eff}|E_0|^2\omega_a, 
\label{eqn:psignal}
\ee
where $Q_{\rm sys}$ is the loaded quality factor, $\kappa$ is a coupling/form factor, and $f_+=b^2/(\omega^2+b^2)$ is a mode mixing factor. We expect $\kappa\sim 1/\epsilon$, however this could be enhanced by $1/k^2$ at small $k\sim \sqrt{\epsilon}m_a$ due to resonant mixing.

The reference power, $P_0$ (Eq.~\ref{eqn:p0}) is enhanced in Eq.~\eqref{eqn:psignal} by two factors: first the quality factor of the AF resonance; second the effective volume can be far larger than $(2\pi/\omega)^3$. The volume amplification arises from the modified dispersion relation: the resonance is tuned by $B_0$, and is independent of the A-TI volume (the resonance scanning requires no precision THz mechanical motion at cryogenic temperatures). The mode mixing factor leads to a small suppression of power, and determines the optimal material via $b$. The coupling factor, $\kappa$, should be optimised in engineering of coatings, geometry, and material $\epsilon$.  

The effective $Q_{\rm sys}$ is due to the electric field enhancement inside the A-TI due to the modified dispersion relation in Eq.~\ref{eqn:driven_system}. We assume $Q_{\rm sys} = 10^5$, using THz AFMR measurements that report $Q\sim 10-100$ at $T\sim4$ K, reducing at lower $T$~\cite{PhysRev.129.1566,PhysRevLett.111.017204,PhysRevB.87.020408,PhysRevLett.100.157204,2017PhRvL.119v7201L}, and scaling $Q\propto T^{-3}$~\cite{PhysRevB.3.961,PhysRevLett.111.017204,PhysRevB.93.014425,PhysRevB.96.180414} down to $\sim$100 mK dilution refrigerator temperatures common in axion searches~\cite{Du:2018uak}. The value of $Q$ and operating temperature will be key drivers in the final choice of material and experimental design. 

The polariton in the A-TI should be optimally coupled to the free space electromagnetic field at the surface for efficient photon measurement, and material losses due to Gilbert damping and phonon production (additional decay channels in Fig.~\ref{fig:feynmans}b) should be of order the photon emission. We absorb into $V_{\rm eff}$ (see Eq.~\ref{eqn:p0}) the relevant form factors, the effect of the A-TI dielectric constant, and any boost factor, $\beta^2$, arising from the geometry~\cite{Millar:2016cjp}. For $V_{\rm eff}=1\text{ cm}^3$, $g_\gamma=10^{-10}\text{ GeV}^{-1}$, $B_0=2$ T ($\omega_a=0.8 \text{ meV}$, $\nu=210$ GHz), and $\kappa=0.01$, the power is $5\times 10^{-22}\text{ W}$: about one photon every 0.3 seconds. 

We use SPD to estimate the measurement sensitivity because, at low temperatures and high frequencies, it is more advantageous than power detection~\cite{2013PhRvD..88c5020L}. While phase-insensitive linear amplifications are fundamentally limited by the standard quantum limit, SPD suffers no strict sensitivity limit. A high confidence detection requires the dark count rate, $\Gamma_d$, of the detector to be smaller than the flux. We use $\Gamma_d=0.001 \text{Hz}$, which has been demonstrated for the quantum dot detector in THz regime at 0.05 K~\cite{KomiyamaQuantumDot}. A wider bandwidth, lower dark count SPD using graphene-based Josephson junction~\cite{2017PhRvP...8b4022W} has the potential to improve significantly the search for heavy dark axions in the future, including our proposal.

We propose to shield backgrounds by placing the entire apparatus in a cryostat, and then measure the baseline photon count at $B_0=0$. Measuring the dependence of the signal on $B_0$ and other features of the theoretical DA lineshape (measured using a bandpass)~\cite{2010PhRvL.104d1301A,Du:2018uak,OHare:2017yze} will allow candidate lines to be distinguished from signal.

The range of axion masses accessible to our technique depends on the scaling of material properties with $B_0$. We take $1\text{ T}<B_0<10\text{ T}$ with stability $\delta B_0=10^{-3}\text {T}$ over the volume, which has been demonstrated~\cite{PENG2000569, Du:2018uak}. For the parameters of \BiFeSe\, given above and setting $\omega_+(k=0)=m_a$ we find $0.7\text{ meV}\leq m_a\leq 3.5\text{ meV}$ (the lower limit is approximately the $B_0=0$ spin wave mass). Other materials with different anisotropy field strengths can cover a wider range of masses.

\begin{figure}
\vspace{-0.2em}\includegraphics[width=1\columnwidth]{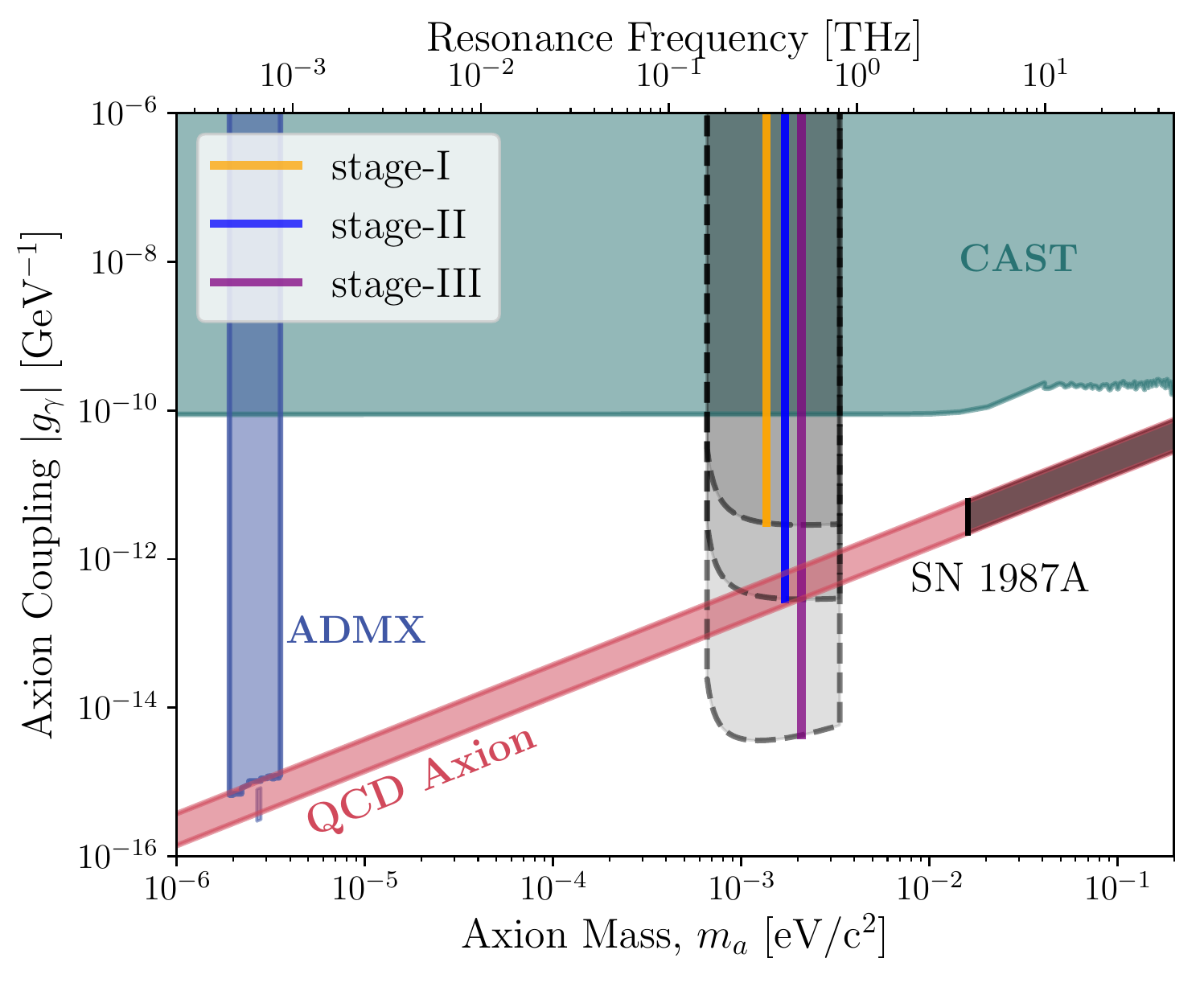}
\caption{Axion parameter space. Vertical lines lines show the projected sensitivity of our proposal using Fe doped Bi$_2$Se$_3$ at $\sim$5T applied field for $10^2$ s integration time with dark count rate $\Gamma_d=0.001 \text{ Hz}$. Staged designs are described in the text. Gray shaded regions assume scanning $1\text{ T}\leq B_0\leq 10\text{ T}$. The KSVZ and DFSZ axion models are shown as the red band. Existing exclusions from ADMX~\cite{2010PhRvL.104d1301A,Du:2018uak}, CAST~\cite{2014PhRvL.112i1302A}, and supernova 1987A~\cite{2008LNP...741...51R} are shown as coloured regions.}  
\label{fig:snr}
\end{figure}

Sensitivity to $g_\gamma$ is computed setting the signal to noise ratio ${\rm SNR}=3$. We take the measurement time on a single frequency $\tau = 10^2\text{ s}$. The full range can be scanned in \Tscans. The volume of any single, high quality, sample of A-TI is limited to be less than \VTI\, to achieve homogeneous doping~\cite{Lostak1997}. The sensitivity is shown in Fig.~\ref{fig:snr} (stage-I). 

Using $N$ A-TI samples, either with a simple tiling and use of lenses, or with coherent addition~\cite{TheMADMAXWorkingGroup:2016hpc}, the gain in $V_{\rm eff}$ can increase linearly with $N$, with wide band response~\cite{Millar:2016cjp}. With $N=100$ (a feasible total number for solid state synthesis~\cite{WEI2015417}), the increased sensitivity is shown in Fig.~\ref{fig:snr} (stage-II).

A further increase in $V_{\rm eff}$ can be achieved by surrounding the A-TI samples with a cavity with a volume, $V_c$. Long wavelength modes of the cavity $E$-field can couple to high frequency AP modes resulting in a TM$_{010}$ type~\cite{Bradley:2003kg} component to the AP, allowing $V_{\rm eff}\approx V_c$ even with a small sample volume. In Fig.~\ref{fig:snr} (stage-III) we show the sensitivity benefit of a $V_{\rm eff}=(0.1 \lambda_{\rm dB} )^3\approx 2000(1\text{ meV}/m_a)^3\text{ cm}^3$. The same stage-III sensitivity could be achieved if technology and investment allowed for fabrication of a very large volume of A-TI.

In summary, we have shown that A-TIs can host dynamical axionic quasiparticles which are resonantly driven in the presence of DAs with mass of order 1 meV and emit THz photons which can be detected using an SPD, allowing A-TIs to detect dark matter. We showed that antiferromagnetic Fe-doped Bi$_2$Se$_3$ satisfies the three Wilczek criteria described earlier, and can be used to realize a DA detector in the \mrange\, range. Fig.~\ref{fig:snr} shows the projected reach of three possible schemes with different effective volumes. Varying the applied $B$ field scans the resonant frequency, giving sensitivity to axion dark matter in a parameter space inaccessible to other methods. Future work on the material characteristics (such as the anisotropy field strength) can allow for a wider range of DA mass detection.  

\acknowledgements{We acknowledge useful discussions with Francesca Day, Joerg Jaeckel, Alexander Millar, Jens Niemeyer, David Tanner, Chris Weber, Willian Witczak-Krempa, and Ariel Zhitnisky.  MA and DJEM are supported by the Alexander von Humboldt Foundation and the German Federal Ministry of Education and Research. KCF acknowledges support from Army Research Office under Cooperative Agreement Number W911NF-17-2-0086. L.\v{S}. acknowledges support from the EU FET Open RIA Grant 766566.}

\bibliographystyle{h-physrev3.bst}

\bibliography{axion_review}

\end{document}